\documentclass[aps,preprint,superscriptaddress]{revtex4}

\usepackage{graphicx}
\usepackage{latexsym}
\usepackage{amsmath}
\usepackage{amssymb}
\usepackage{amsfonts}
\usepackage{color}

\begin{document}

\title{Type-1.5 Superconductors}

\author{V. V. Moshchalkov}
\affiliation{INPAC-Institute for Nanoscale Physics and Chemistry,
Nanoscale Superconductivity and Magnetism $\&$ Pulsed Fields
Group, K. U. Leuven Celestijnenlaan 200 D, B-3001 Leuven,
Belgium.}

\author{M. Menghini}
\affiliation{INPAC-Institute for Nanoscale Physics and Chemistry,
Nanoscale Superconductivity and Magnetism $\&$ Pulsed Fields
Group, K. U. Leuven Celestijnenlaan 200 D, B-3001 Leuven,
Belgium.}

\author{T. Nishio}
\affiliation{INPAC-Institute for Nanoscale Physics and Chemistry,
Nanoscale Superconductivity and Magnetism $\&$ Pulsed Fields
Group, K. U. Leuven Celestijnenlaan 200 D, B-3001 Leuven,
Belgium.}

\author{Q.H. Chen}
\affiliation{INPAC-Institute for Nanoscale Physics and Chemistry,
Nanoscale Superconductivity and Magnetism $\&$ Pulsed Fields
Group, K. U. Leuven Celestijnenlaan 200 D, B-3001 Leuven,
Belgium.}

\author{A. V. Silhanek}
\affiliation{INPAC-Institute for Nanoscale Physics and Chemistry,
Nanoscale Superconductivity and Magnetism $\&$ Pulsed Fields
Group, K. U. Leuven Celestijnenlaan 200 D, B-3001 Leuven,
Belgium.}

\author{ V.H. Dao}
\affiliation{INPAC-Institute for Nanoscale Physics and Chemistry,
Nanoscale Superconductivity and Magnetism $\&$ Pulsed Fields
Group, K. U. Leuven Celestijnenlaan 200 D, B-3001 Leuven,
Belgium.}

\author{L.F. Chibotaru}
\affiliation{INPAC-Institute for Nanoscale Physics and Chemistry,
Nanoscale Superconductivity and Magnetism $\&$ Pulsed Fields
Group, K. U. Leuven Celestijnenlaan 200 D, B-3001 Leuven,
Belgium.}

\author{N. D. Zhigadlo}
\affiliation{Solid State Physics Laboratory ETH-Z\"{u}rich,
Switzerland.}

\author{J. Karpinski}
\affiliation{Solid State Physics Laboratory ETH-Z\"{u}rich,
Switzerland.}

%\thanks{}

\date{\today}
\begin{abstract}
We demonstrate the existence of a novel superconducting state in
high quality two-component MgB$_2$ single crystalline
superconductors where a unique combination of both type-1
($\lambda_1/ \xi_1<1/ \sqrt2$) and type-2
($\lambda_2/\xi_2>1/\sqrt2$) superconductor conditions is realized
for the two components of the order parameter. This condition
leads to a vortex-vortex interaction attractive at long distances
and repulsive at short distances, which stabilizes unconventional
{\it stripe- and gossamer-like vortex patterns} that we have
visualized in this {\it type-1.5} superconductor using Bitter
decoration and also reproduced in numerical simulations.

\end{abstract}

%\keywords{}

\maketitle

Vortices in superconductors are characterized by a normal core
with size $\xi$ (coherence length) and supercurrents flowing over
a distance $\lambda$ (penetration depth). If two vortices are
generated in a type-1 superconductor the normal cores would
overlap first, due to the larger value of $\xi$ with respect to
$\lambda$, thus leading to a gain in the condensation energy and,
consequently, to vortex-vortex attraction\cite{kramer,brandt}. Two
vortices in a type-2 material, however, would have first their
supercurrents overlapping, in view of the bigger $\lambda$,
leading to vortex-vortex repulsion. An attractive vortex-vortex
interaction results in the formation of macroscopic normal domains
in the intermediate state\cite{huebener}, while vortex-vortex
repulsion leads to the appearance of the Abrikosov
lattice.\cite{abrikosov}

The recent discovery of the two-component superconductor
MgB$_2$\cite{nagamatsu}, with two weakly coupled coexisting order
parameters, $\Psi_1 =\Psi_{\pi}$ and $\Psi_2 =\Psi_{\sigma}$, has
opened remarkable new possibilities both for fundamental research
and applications. Among the new research topics we find
semi-Meissner state\cite{babaev1}, the violation of the London law
and Onsager-Feynman quantization\cite{babaev2}, non-composite
vortices \cite{chibotaru}, intrinsic Josephson effect with
low-energy interband Josephson plasmons \cite{blumberg}, mapping
physics of two-component superconductors on two-condensate Bose
systems\cite{babaev3}, superfluidity in liquid metallic hydrogen
\cite{babaev4}, etc. The two-component character of
MgB$_2$\cite{bouquet,szabo,giubileo,souma} is related with two
different types of electronic bondings,  $\pi$ and $\sigma$,
giving rise to two different superconducting gaps with energies
$\Delta_{\pi}(0)= 2.2\,$meV \cite{rubio,eskildsen} and
$\Delta_{\sigma}(0)=7.1\,$meV \cite{golubov1,choi,iavarone},
respectively. Using the BCS expression $\xi(0)=\hbar v_F/ \pi
\Delta(0)$, where $v_F$ is the Fermi velocity (5.35$\times
10^{5}~$m/s for the $\pi$-band and 4.40$\times 10^{5}~$m/s for the
$\sigma$-band)\cite{brinkman} for the $\pi$- and $\sigma$- bands
separately, we obtain the two coherence lengths:
$\xi_{\pi}(0)=51\,$nm and $\xi_{\sigma}(0)=13\,$nm. The calculated
$\xi_{\pi}(0)$ value is in agreement with $\xi_{\pi}(0)=49.6\pm
0.9\,$nm obtained from the fit of the vortex profile measured by
scanning tunneling spectroscopy (STS) \cite{eskildsen}. The London
penetration depths $\lambda_{\pi}$ and $\lambda_{\sigma}$ can be
found from the respective plasma frequencies $\omega_{p\pi}$ and
$\omega_{p\sigma}$ \cite{mazin}: $\lambda_{\pi}(0)=33.6\,$nm and
$\lambda_{\sigma}(0)=47.8\,$nm.
 As a result, at least in
the clean limit, the  $\pi$- and $\sigma$- components of the order
parameter in MgB$_2$ are in different regimes:
$\kappa_{\pi}=\lambda_{\pi}(0)/\xi_{\pi}(0)=0.66\pm0.02
<1/\sqrt{2}$=0.707 (type-1) (assuming $\lambda_{\pi}(0)=33.6\,\pm
0.9~$nm) and
$\kappa_{\sigma}=\lambda_{\sigma}(0)/\xi_{\sigma}(0)=3.68>1/\sqrt{2}$
(type-2). However, when MgB$_2$ is doped with impurities, both
components of the order parameter (due to an increase of $\lambda$
and a decrease of $\xi$) will fall inevitably into the type-2
regime. Therefore, clean MgB$_2$ represents an excellent candidate
to search for a new type of superconductivity, neither of type-1
 nor type-2 character, which we coined as $\emph{type-1.5
superconductivity}$. In type-1.5 superconductors the vortex-vortex
interaction is the result of the competition between short-range
repulsion and long-range attraction and it is expected that
vortices could form unusual patterns at low applied
fields.\cite{babaev1}

In this Letter, we present experimental observation of vortex
patterns at low vortex densities in high quality MgB$_{2}$ single
crystals. The vortex patterns are compared with the results of
molecular dynamics simulations based on a two-gap Ginzburg-Landau
(GL) theory  which results in peculiar equilibrium vortex
structures such as gossamer-like vortex patterns and vortex
stripes.

The MgB$_2$ single crystals were grown using a high pressure
method as described elsewhere \cite{karpinski}. The critical
temperature of the samples is typically $38.6\,$K as determined
from the ac-susceptibility response at zero field.  The
temperature dependence of the lower and upper critical fields
$H_{c1}(T)$ and $H_{c2}(T)$ were determined by means of
magnetization measurements \cite{perkins, moshchalkov} and by
ac-susceptibility measurements, respectively. From the
extrapolated value $H_{c2}(0)= 5.1\,$T we obtain $\xi_{ab}(0)=
8.0\,$nm close to the value found from BCS theory for
$\xi_{\sigma}(0)$. In order to estimate the penetration depth of
the $\sigma$-band we use the theoretical expression for the field
at which the first vortex will penetrate a two component
superconductor [Eq. (4) in Ref. \cite{babaev1}]. Considering
$\lambda_{\pi}(0)=33.6\,$nm \cite{mazin} for the type-1 component
and taking into account $\xi_{ab}(0)=\xi_{\sigma}(0)=8.0\,$nm and
$H_{c1}(0)=0.241\,$T from our measurements we obtain
$\lambda_{ab}=\lambda_{\sigma}=38.2\,$nm for our samples.

Coexisting two interpenetrating weakly coupled order parameters is
a new quantum state with superflow of the two components through
each other without any friction. A vortex-vortex interaction is
derived from the GL theory by numerically minimizing the free
energy of two vortices with a variational procedure\cite{jacobs}.
We use the two-band GL functional:

\begin{equation}
F_{GL}[\Psi_{\sigma},\Psi_{\pi},\textbf{A}]
=F_{\sigma}+F_{\pi}+\int
d^{3}r[\frac{E_{\gamma}}{2}(\Psi_{\sigma}^{\ast}\Psi_{\pi}+\Psi_{\pi}^{\ast}\Psi_{\sigma})+\frac{1}{2\mu_{0}}(\nabla\times\textbf{A})^2]
\end{equation}

where
\begin{equation}
F_{\alpha}=\int
d^{3}r[2E_{c\alpha}|\Psi_{\alpha}|^{2}+|E_{c\alpha}||\Psi_{\alpha}|^{4}+\frac{\Phi_{0}}{2\pi}\sqrt{\frac{|E_{c\alpha}|}{\mu_{0}\kappa_{\alpha}^{3}}}|(-i\nabla+\frac{2\pi}{\Phi_{0}}\textbf{A})
\Psi_{\alpha}|^{2}]
\end{equation}

where $\alpha=\sigma,\pi$ and $\Phi_{0}$ is the flux quantum. The
estimations for the intrinsic condensation energies $E_{c\alpha}$
and the coupling energy $E_{\gamma}$ are taken from Ref.
\cite{eisterer} and we use the values of $\kappa_{\sigma}$ and
$\kappa_{\pi}$ obtained for our samples. The result of the
minimization shows that the interaction between vortices is
short-range repulsive and weakly long-range attractive, similarly
to Ref. \cite{babaev1}. We model a system of overdamped vortices
by molecular dynamics simulations. The equation of motion for a
vortex $i$ is
$\textbf{F}_{i}=\textbf{F}_{i}^{vv}+\textbf{F}_{i}^{T}=\eta
\textbf{v}_{i}$, where $\textbf{F}_{i}^{vv}$ represents the
vortex-vortex interaction and $\textbf{F}_{i}^{T}$ the thermal
stochastic force satisfying $\langle
\textbf{F}_{i}^{T}(t)\rangle=0$ and $\langle
\textbf{F}_{i}^{T}(t)\textbf{F}_{j}^{T}(t')\rangle=2\eta\delta_{i,j}\delta(t-t')k_{B}
T$. $\eta$(=$\Phi_{0}H_{c2}/\rho_{n}$) is the viscosity,
$H_{c2}$=5.1 T is the upper critical field and $\rho_{n}$=0.7
$\mu\Omega$$\cdot$cm \cite{eisterer} is the normal state
resistivity. The system size used in our simulations is
2000$\times$2000$\lambda^{2}(0)$. Two systems with number of
vortices $N_{v}=150$ and 400 are initially prepared in a high
temperature molten state and then slowly annealed to $T=4.2\,$K
with 1000 temperature steps. We let the system stabilize during
2000 time steps in each step of temperature. In order to compare
the vortex structure of a type-1.5  with a conventional type-2
superconductor we also carry out this simulation for NbSe$_{2}$
using the superconducting parameters from Ref. \cite{higgins}.

Bitter decoration experiments on MgB$_2$ single crystals were
performed at $4.2\,$K after cooling down the sample from above
$T_c$ in the presence of an applied field perpendicular to the
sample surface (field cooling, FC). In this way, a homogeneous
vortex distribution all over the sample is expected. Details about
the technique can be found elsewhere \cite{fasano}. A Bitter
decoration image at $H = 1\,$Oe shows clear evidence of an
inhomogeneous distribution of vortices [Fig.\,\ref{figure1}(a)]
reminding gossamer patterns: local groups of vortices with
intervortex distances shorter than the average vortex distance
($2\Phi_0/\sqrt{3}B)^{1/2}\sim5 \mu\,$m are separated by randomly
distributed vortex voids with size of few micrometers. This is in
striking contrast with the conventional homogenous vortex pattern
formed in NbSe$_2$ single crystals [see Fig.\,\ref{figure1}(b)].
The observed clustering of vortices in MgB$_2$ samples is
consistent with the theoretical modeling made in
Ref.\,\cite{babaev1} for a two-component superconductor in the
semi-Meissner state. In Fig.\,\ref{figure2}(a) the vortex
positions in a selected region of the Bitter decoration image
shown in Fig.\,\ref{figure1}(a) are indicated as white dots, while
in Fig.\,\ref{figure2}(b) the results from the numerical
simulations for a two-component superconductor are shown. We also
calculate the vortex configuration for a reference conventional
type-2 superconductor. The obtained vortex structure, considering
$\lambda=69\,$nm and $\xi=7.7\,$nm \cite{higgins}, is similar to
the one observed in NbSe$_2$ samples [see Figs.\,\ref{figure2}(c)
and (d)].

In order to characterize the inhomogeneity of the vortex structure
(VS)  we calculate the distribution of first neighbor distance,
$P_a$ [Figs.\,\ref{figure2}(e) and (f)]. The first neighbor
distance, $a$, is calculated by means of the Delaunay
triangulation of the vortex structure. For the  VS of NbSe$_2$,
the distribution is Gaussian with a relative standard deviation
$\delta=SD/<a>=0.224$, where SD is the standard deviation of the
Gaussian fit to the experimental data and $<a>$ is the average
first neighbor distance. On the other hand, in MgB$_2$ samples
$P_a$ is  quite broad as a consequence of the inhomogeneous
arrangement of vortices and has additional  peaks at distances
shorter and longer than the most probable separation [see the red
and green arrows in Fig.\,\ref{figure2}(e)]. The distribution of
the first neighbor separation of the vortex structure obtained by
simulations of a two-component material has also a three-peak
structure which is fairly broader when compared to the one
obtained in the case of a one-component conventional type-2
superconductor [Fig.\,\ref{figure2}(f)]. The peaks at short
distances in Figs.\,\ref{figure2}(e) and (f) correspond to an
average minimum separation between vortices.

Figure\,\ref{figure3}(a) shows a Bitter decoration image of the VS
in MgB$_2$ crystals at $H=5\,$Oe and $T=4.2\,$K. The vortex
distribution appears also to be inhomogeneous at this field but in
a rather different manner than in the experiment described above.
In some regions of the sample, vortices agglomerate forming
stripes while voids are formed on considerably larger
areas.\cite{authors}  The present experimental results seem to be
in contradiction with previous reports of the vortex structure at
low fields in MgB$_2$ samples \cite{vinnikov}. However, it is
important to note that in Ref.\,\onlinecite{vinnikov} the authors
show an image of the VS at $4\,$Oe in a very small region of the
sample (approximately $10\times10\,\mu$m$^2$). Therefore, it is
not possible to determine whether the vortex distribution in the
samples studied in Ref. \cite{vinnikov} is uniform all over the
sample or not.

Although the vortex stripe pattern is rather disordered it is
still possible to determine an average direction of vortex stripes
as the one defined by the dashed yellow lines in
Fig.\,\ref{figure3}(a). Considering this definition, we calculate
the vortex density in lines parallel to the vortex stripes as a
function of the distance measured along the direction of the
yellow arrow in Fig.\,\ref{figure3}(a). In the inset of
Fig.\,\ref{figure3}(b) we plot the linear vortex density
normalized by the average value for both the MgB$_2$ and
NbSe$_{2}$ VS at $5\,$Oe. In the case of MgB$_2$, fluctuations of
the vortex density of the order of $50\%$ are observed. A similar
calculation along lines perpendicular to the stripes shows that
the standard deviation of the mean value is of the order of
$30\%$. The large fluctuations of the vortex density in MgB$_2$
are in contrast to what is observed in NbSe$_{2}$ crystals where
the standard deviation of the vortex density is approximately
$1\%$ of the average value. A remarkable similarity is found
between the experiments and simulations at still low density but
higher than the one shown in Fig.\,\ref{figure2}. Disordered
vortex stripes are formed in the two-component superconductor
while a homogeneous distribution is apparent in the case of a
conventional type-2 material [Figs.\,\ref{figure3}(c) and (d),
respectively]. Consistently, the vortex density is seen to
fluctuate in the direction perpendicular to the vortex stripes in
the two-component material, as shown in the inset of
Fig.\,\ref{figure3}(d).

Composition analysis (via an electron microprobe in a field
emission scanning electron microscope) in an area across the
stripes [in the direction of the white arrow in Fig. 3(a)] shows
no significant variations in Mg or B content, thus ruling out the
possibility to attribute the stripe formation to inhomogeneous
surface pinning distribution. There is also no observed
correlation between the vortex positions and localization of
microdefects.

Bitter decoration images at $H=0.5\,$Oe also show a very
inhomogeneous vortex distribution (not shown). On the other hand,
at $H=10\,$Oe the VS in MgB$_2$ samples is similar to the one in
NbSe$_{2}$ crystals indicating that this novel superconducting
phase in two-component type-1.5 superconductors is only accessible
at very low applied fields as predicted in Ref. \cite{babaev1}.

In conclusion, the $\emph{type-1.5}$ superconductivity is a
totally new state which combines two regimes (type-1 and type-2)
in the same single material (clean MgB$_2$ and possibly also other
two-gap materials such as Ba$_{0.6}$K$_{0.4}$Fe$_2$As$_2$
\cite{ding}). The vortex matter in type-1.5 superconductors
behaves in an extremely unusual way. The combination of the
vortex-vortex repulsion and attraction in the same material leads
to the appearance of novel vortex patterns: gossamer-like vortex
arrays and vortex stripes. Both novel patterns have been directly
visualized by Bitter decorations on high quality single crystals.
Moreover, analytical modeling of exotic vortex-vortex interaction
(attractive mixed with repulsive) and extensive molecular dynamic
simulations are in good agreement with our experimental data.

We would like to thank P. L'Ho\"{e}st from the MTM Department in
K. U. Leuven for performing the electron probe microanalysis
experiments on the MgB$_2$ crystals. This work was supported by
Methusalem Funding of the Flemish government, FWO-Vlaanderen, and
the Belgian Inter-University Attraction Poles IAP Programmes.

%% figure ChC_layout%%

\newpage
%% figure ChC_layout%%
\begin{figure}
\centering
\caption{Magnetic decoration images of the vortex structure at $T$
= 4.2 K and $H$ = 1 Oe. The vortices in the MgB$_2$ single crystal
(a) are distributed inhomogeneously while in the NbSe$_{2}$ single
crystal (b) they are arranged forming a disordered Abrikosov
lattice. The scale bars in the images correspond to 10 $\mu$m.
Notice that the density of vortices in the decoration experiments
represents the internal field $B$ rather than the applied field
$H$. This leads to different number of decorated vortices for
NbSe$_{2}$ and MgB$_{2}$, even at the same applied field.}
  \label{figure1}
\end{figure}

%% figure ChC_layout%%
\begin{figure}
\centering
\caption{(a) Experimental vortex locations in a selected part of
the image shown in Fig. 1(a). The vortex configuration resulting
from the numerical simulations in a two-component superconductor
at low density is shown in (b) evidencing an inhomogeneous spatial
distribution of vortices. In both cases, the regions enclosed by
the dashed white line indicate voids of vortices caused by the
inhomogeneous distribution. In (c) the vortex pattern obtained by
a magnetic decoration of the NbSe$_{2}$ crystal at 1 Oe is shown
and (d) corresponds to the vortex pattern obtained by a numerical
simulation of a type-2 superconductor. The white scale bars
correspond to 10 $\mu$m. (e) and (f) display the distribution of
first neighbor distance, $P_{a}$, of the experimental and
theoretical vortex structures, respectively. The distributions
obtained for the simulations of a type-2 material and for the
NbSe$_{2}$ vortex structure are Gaussian. In the case of MgB$_2$
$P_{a}$ shows additional peaks at distances shorter and longer
than the most probable separation (see the red and green arrows).
Pair of vortices separated at the distances where the additional
peaks are located are highlighted in (a) and (b) by red and green
circles. The light blue circles correspond to pair of vortices
separated by the most probable separation.}
  \label{figure2}
\end{figure}

%% figure ChC_layout%%
\begin{figure}
\centering
\caption{(a) Magnetic decoration image showing a stripe-like
vortex pattern in the MgB$_2$ single crystal at $H=5\,$Oe.
(b)Disordered Abrikosov lattice with a homogeneous vortex density
obtained at $H=5\,$Oe in the NbSe$_2$ sample. The formation of
vortex stripes is also observed by numerical simulations of a
two-component type-1.5 superconductor (c) in contrast to a
homogeneous vortex distribution in a type-2 superconductor at the
same vortex density (d). The scale bars in the images correspond
to 10 $\mu$m. Vortex density along lines parallel to the vortex
stripe direction (yellow dashed lines in (a)) for MgB$_2$ and
NbSe$_{2}$ vortex structures. The variation of the vortex density
as a function of the distance measured along the direction
perpendicular to the stripes (yellow arrows in (a) and (b)) is
shown in the insert of (b). The curves are normalized by their
respective average density.  The results of a similar calculation
performed on the simulated vortex structures are shown in the
inset of (d).}
  \label{figure3}
\end{figure}


\begin{thebibliography}{30}

\bibitem{kramer} L. Kramer, Phys. Rev. B \textbf{3}, 3821 (1971).

\bibitem{brandt} E. H. Brandt Phys. Rev. B \textbf{34} 6514
(1986).

\bibitem{huebener}R. P. Huebener, {\it Magnetic Flux Structures of
Superconductors}, (Springer-Verlag, New York, 1990).

\bibitem{abrikosov}A. A. Abrikosov, Zh. Eksp. Teor. Fiz. {\bf 32}, 1442 (1957).

\bibitem{nagamatsu} J. Nagamatsu et al., Nature \textbf{410}, 63 (2001).

\bibitem{babaev1} E. Babaev and M. Speight, Phys. Rev. B \textbf{72}, 180502(R) (2005).

\bibitem{babaev2} E. Babaev and N. W. Ashcroft, Nature Physics \textbf{3}, 530 (2007).

\bibitem{chibotaru} L. F. Chibotaru, V. H. Dao, and A. Ceulemans, Europhys. Lett. \textbf{78}, 47001  (2007).

\bibitem{blumberg} G. Blumberg, et al., Phys. Rev. Lett. \textbf{99}, 227002 (2007).

\bibitem{babaev3} E. Babaev, L. D. Faddeev, and A. J. Niemi, Phys. Rev. B \textbf{65}, 100512(R) (2002).

\bibitem{babaev4} E. Babaev, A. Sudb{\o}, and N. W. Ashcroft, Nature \textbf{431}, 666 (2004).

\bibitem{bouquet} F. Bouquet, R. A. Fisher, N. E. Phillips, D. G. Hinks, and J. D. Jorgensen , Phys. Rev. Lett. \textbf{87}, 047001 (2001).

\bibitem{szabo} P. Szabo et al., Phys. Rev. Lett. \textbf{87},
137005 (2001).

\bibitem{giubileo} F. Giubileo et al., Phys. Rev. Lett. \textbf{87}, 177008 (2001).

\bibitem{souma} S. Souma et al., Nature \textbf{423}, 65 (2003).

\bibitem{rubio} G. Rubio-Bollinger, H. Suderow, and S. Vieira, Phys. Rev. Lett. \textbf{86}, 5582-5584 (2001).

\bibitem{eskildsen} M. R. Eskildsen et al., Phys. Rev. Lett. \textbf{89}, 187003 (2002).

\bibitem{golubov1} A. A. Golubov et al., J. Phys. Condens. Matter \textbf{14}, 1353 (2002).

\bibitem{choi} H. J. Choi et al., Nature (London) \textbf{418}, 758 (2002).

\bibitem{iavarone} M. Iavarone et al., Phys. Rev. Lett. \textbf{89}, 187002 (2002).

\bibitem{brinkman} A. Brinkman et al. Phys. Rev. B \textbf{65},
180517(R) (2002).

\bibitem{mazin} I. I. Mazin et al., Phys. Rev. Lett. \textbf{89}, 107002 (2002).


\bibitem{karpinski} J. Karpinski et al., Supercond. Sci. Technol. \textbf{16}, 221 (2003).

\bibitem{moshchalkov} V. V. Moshchalkov et al., Physica C \textbf{175}, 407
(1991).
\bibitem{perkins} G. K. Perkins et al., Supercond. Sci. Technol.
\textbf{15} 1156 (2002).

\bibitem{jacobs} L. Jacobs and C. Rebbi, Phys. Rev. B \textbf{19}, 4486 (1979).

\bibitem{eisterer}  M. Eisterer, Supercond. Sci. Technol. \textbf{20}, R47 (2007).

\bibitem{higgins} M. J. Higgins and S. Bhattacharya, Physica C \textbf{257}, 232 (1996).

\bibitem{fasano} Y. Fasano and M. Menghini, Supercond. Sci. and Tech. \textbf{21}, 023001 (2008).

\bibitem{authors}It is important to note that this agglomeration in the form of
vortex stripes is not related to the presence of steps or surface
defects in the crystal. In fact, there are regions where steps can
be observed (not shown) but at those locations the vortex
configuration revealed by Bitter decoration is unaffected by them.

\bibitem{vinnikov} L. Y. Vinnikov et al., Phys. Rev. B \textbf{67}, 092512 (2003).

\bibitem{ding} H. Ding et al., Europhys. Lett. \textbf{83}, 47001 (2008).





\end{thebibliography}
\end{document}